\documentclass[slac_one]{revtex4}
\usepackage{graphicx}
\usepackage{fancyhdr}
\newcommand{\epem}     {\ensuremath{\mathrm{e^+e^-}}}
\newcommand{\roots}    {\ensuremath{\sqrt{s}}}
\newcommand{\znull}    {\ensuremath{\mathrm{Z^0}}}
\newcommand{\mz}       {\ensuremath{m_{\znull}}}
\newcommand{\mw}       {\ensuremath{m_{\mathrm{W}}}}
\newcommand{\wplus}    {\ensuremath{\mathrm{W^+}}}
\newcommand{\wminus}   {\ensuremath{\mathrm{W^-}}}
\newcommand{\wpwm}     {\ensuremath{\mathrm{W^+W^-}}}
\newcommand{\order}[1] {\mbox{\ensuremath{{\cal O}(#1)}}}
\newcommand{\momone}[1] {\mbox{\ensuremath{\langle#1\rangle}}}
\newcommand{\chisqd}   {\ensuremath{\chi^2/\mathrm{d.o.f.}}}

\newcommand{\expt}   {\ensuremath{\mathrm{(exp.)}}}

\newcommand{\theo}   {\ensuremath{\mathrm{(theo.)}}}
\newcommand{\ddel}   {\ensuremath{\mathrm{d}}}

\newcommand{\bt}     {\ensuremath{B_T}}
\newcommand{\bw}     {\ensuremath{B_W}}
\newcommand{\bn}     {\ensuremath{B_N}}
\newcommand{\mh}     {\ensuremath{M_H}}
\newcommand{\ml}     {\ensuremath{M_L}}
\newcommand{\thr}    {\ensuremath{1-T}}

\newcommand{\tmin}   {\ensuremath{T_{\mathrm{min.}}}}
\newcommand{\cp}     {\ensuremath{C}}
\newcommand{\dpar}   {\ensuremath{D}}
\newcommand{\ytwothree} {\ensuremath{y_{23}}}
\newcommand{\ytwothreed} {\ensuremath{\ytwothree^{\mathrm{D}}}}
\newcommand{\ythreefour} {\ensuremath{y_{34}}}
\newcommand{\ythreefourd} {\ensuremath{\ythreefour^{\mathrm{D}}}}

\newcommand{\as}     {\ensuremath{\alpha_{\mathrm{S}}}}
\newcommand{\asnp}   {\ensuremath{\as^{\mathrm{NP}}}}
\newcommand{\asmz}   {\ensuremath{\as(\mz)}}
\newcommand{\nf}     {\ensuremath{n_f}}

\newcommand{\atau}   {\ensuremath{\alpha_{\tau}}}

\newcommand{\mui}    {\ensuremath{\mu_{\mathrm{I}}}}
\newcommand{\anull}  {\ensuremath{\alpha_0}}
\newcommand{\aone}   {\ensuremath{\alpha_1}}

\newcommand{\anulltwo} {\ensuremath{\anull(2~\mathrm{GeV})}}
\newcommand{\aonetwo} {\ensuremath{\aone(2~\mathrm{GeV})}}
\newcommand{\fpt}    {\ensuremath{F_{\mathrm{PT}}}}
\newcommand{\cy}     {\ensuremath{c_{\mathrm{y}}}}

\begin{document}

\title{Power Corrections in Electron-Positron Annihilation: 
       Experimental Review}

%

\author{S. Kluth}
\affiliation{Max-Planck-Institut f\"ur Physik, F\"ohringer Ring 6, D-80805 Munich, Germany}

\begin{abstract}

Experimental studies of power corrections with \epem\ data are
reviewed.  An overview of the available data for jet and event shape
observables is given and recent analyses based on the
Dokshitzer-Marchesini-Webber (DMW) model of power corrections are
summarised.  The studies involve both distributions of the observables
and their mean values.  The agreement between perturbative QCD
combined with DMW power corrections and the data is generally good,
and the few exceptions are discussed.  The use of low energy data sets
highlights deficiencies in the existing calculations for some
observables.  A study of the finiteness of the physical strong
coupling at low energies using hadronic $\tau$ decays is shown.

\end{abstract}

\maketitle

\thispagestyle{fancy}

\section{INTRODUCTION}

Data from hadron production in \epem\ annihilation experiments has
been one of the main sources of experimental information for the study
of non-perturbative QCD of hadron jets, see
e.g.~\cite{kluth06,dasgupta03,kluth01a,kluth01b}.  Compared to other
processes where power corrections have been studied such as hadron
production in deep inelastic lepton-nucleon scattering~\cite{dasgupta03} 
there are two main advantages:
\begin{description}

\item[Leptonic initial state] 

  The incoming electron and positron are
  leptons and thus there is no interference between initial state and
  the hadronic final state.  This makes the interpretation of of the
  final state in terms of QCD much more robust.

\item[Large range of energies] 

  The currently available data from
  \epem\ annihilation to hadrons span a range of centre-of-mass (cms)
  energies of more than an order of magnitude.  Since many effects in
  QCD depend on the energy scale comparing data sets recorded at
  different cms energies yields strong tests of the theory.

\end{description}

From reconstructed hadronic final states jet production or event shape
observables are calculated, see e.q.~\cite{dasgupta03,kluth06}.  As an
example the observable Thrust $T$ is defined by $T=\max_{\vec{n}}
(\sum_i |\vec{n}\cdot\vec{p}_i|)/(\sum_i |\vec{p}_i|)$ where
$\vec{p}_i$ is the 3-momentum of particle $i$ and $|\vec{n}|=1$.  A
value of $T=1$ corresponds to an ideal 2-jet event, events with three
final state objects have $1>T>2/3$ while completely spherical events
with many final state particles can approach $T=0.5$.  Event shape
observables $y$ are conventionally defined to have $y=0$ for ideal
two-jet configurations and thus the Thrust is often analysed in terms
of~\thr.  The observables can be classified as 3-jet or 4-jet
observables depending on whether $y>0$ occurs in the cms for final
states with $\ge3$ or $\ge4$ objects.

\subsection{Experiments}

The energy of the electron and electron beams in most \epem\
annihilation collider experiments is equal due to the simplified
construction and operation of a circular accelerator with only one
beam line.  As a consequence the cms of hadronic final states
produced in \epem\ annihilation coincides with the laboratory
system in the absence of hard initial state radiation (ISR).

\begin{figure}[htb!]
\includegraphics[width=0.5\textwidth, angle=-90]{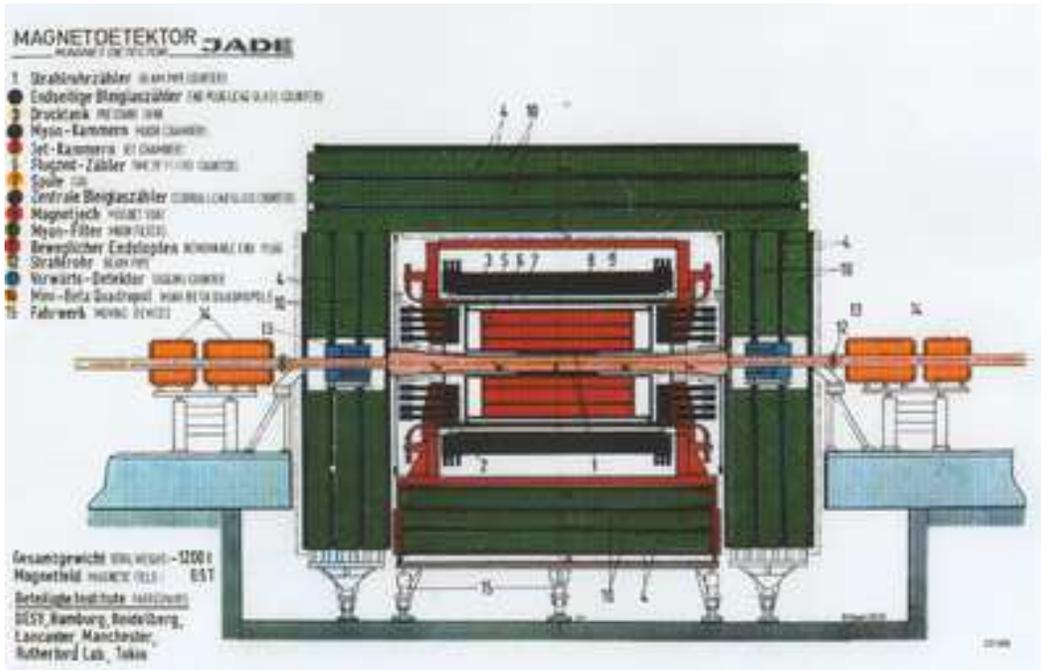}
\caption[ bla ]{ Drawing of the JADE experiment.  The electron
and positron beam come from the left and right and collide
in the centre of the experiment. }
\label{fig_jade}
\end{figure}

A typical \epem\ annihilation experiment is shown in
figure~\ref{fig_jade}.  The JADE detector (for details
see~\cite{naroska87}) was operated at the PETRA
accelerator~\cite{petra80} from 1979 to 1986 at cms energies
$12<\roots<46.8$~GeV.  The design of the experiment is similar to the
more recent LEP experiments.  The interaction region, where the
electron and positron beams collide, is surrounded by gaseous tracking
detectors (Jet Chamber) inside a solenoidal magnetic field of 0.48~T.
The tracking detectors are inside a calorimeter consisting of 2712
lead glass blocks to measure the energy of charged and neutral
particles.  Tracking and calorimeter efficiency is good in the region
$|\cos\theta|<0.97$ were $\theta$ is the angle w.r.t. the beam axis.
Therefore most \epem\ annihilation events are fully contained in
the detector and complete reconstruction with high efficiency and
good precision is possible.  This reduces the size of experimental
corrections for acceptance and resolution and thus the corresponding
experimental uncertainties.

\subsection{Overview of data}

\begin{figure}[htb!]
\begin{tabular}{cc}
\includegraphics[width=0.49\textwidth,clip]{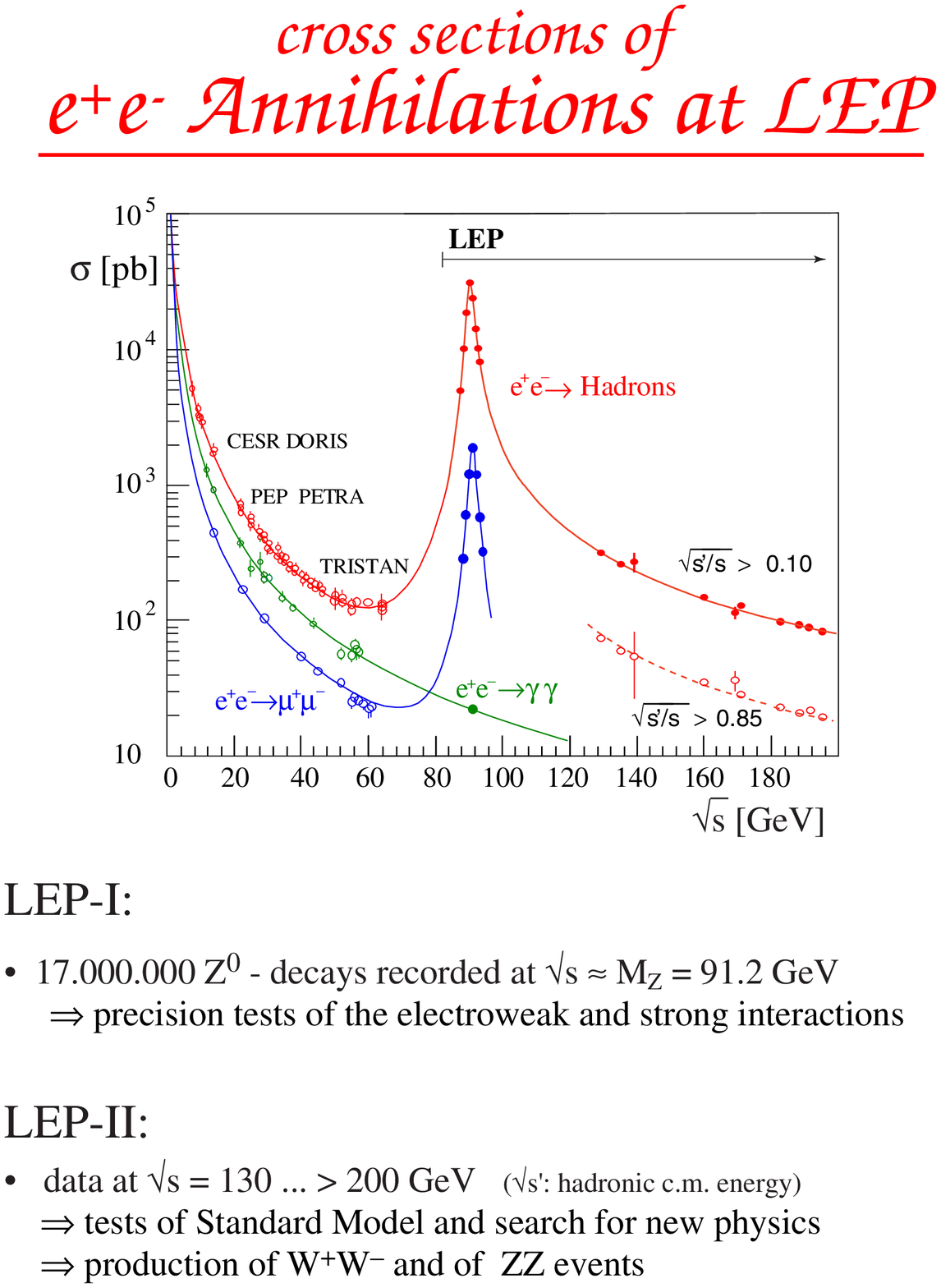} &
\includegraphics[width=0.49\textwidth,clip]{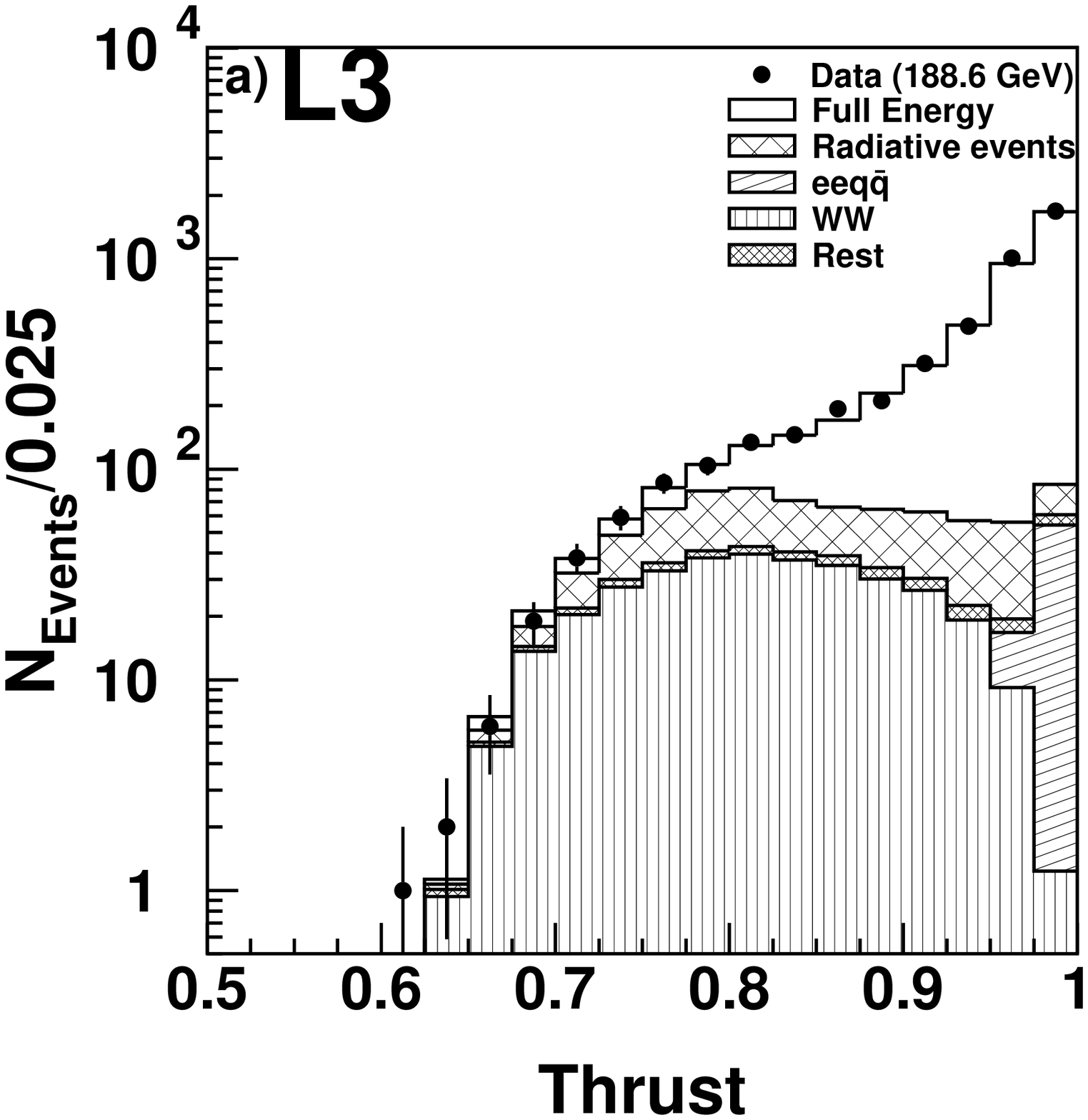} \\
\end{tabular}
\caption[ bla ]{ (left) Cross sections of several \epem\ annihilation
processes as indicated~\cite{bethke00b}.  The lines represent
theoretical predictions.  The accelerators where the data where
recorded are also shown.  The variable $s'$ refers to reduced cms
energy after ISR.  (right) Distribution of event shape observable
Thrust $T$ measured at $\roots\simeq 189$~GeV~\cite{l3290}.  The lines
and hatched regions represent the predictions from Monte Carlo
simulations of signal and background. }
\label{fig_data}
\end{figure}

The experimental conditions vary significantly between the various cms
energies of the data sets.  Figure~\ref{fig_data}
(left)~\cite{bethke00b} shows the measured cross sections of several
\epem\ annihilation processes as a function of the cms energy \roots.
The $1/s$ dependence as expected from QED at low \roots\ is clearly
visible.  At the \znull\ peak around $\roots=91$~GeV the cross sections
increase by a factor of about 100 compared to their lowest
values at $\roots\simeq 60$~GeV or $\roots>130$~GeV.  

At $\roots<\mz$ background from two-photon interactions with hadronic
final states ($\epem\rightarrow\gamma\gamma$), from production of
$\tau$-lepton pairs decaying to hadrons and from hadronic final states
with ISR and thus reduced cms is significant.  Requiring more than
four tracks of charged particles (events with exactly four tracks not
in a 1- vs 3-prong configuration may also be
accepted~\cite{OPALPR299}) reduces the backgrounds from $\tau$-pairs
and two-photon interactions to negligible levels.  The data collected on
or near the \znull\ peak are essentially free of backgrounds due to
the high energy and large cross section.  ISR effects are also
suppressed.  The high energy samples taken by the LEP experiments
during the LEP~2 phase have again backgrounds from two-photon
interactions, ISR and for $\roots>2\mw\simeq161$~GeV from production
of four-fermion final states dominated by \wpwm\ pairs decaying to
hadrons.

As a representative example figure~\ref{fig_data} (right) shows the
distribution of the event shape observable Thrust $T$ measured by L3
at $\roots\simeq 189$~GeV~\cite{l3290}.  The simulation of signal and
background samples shown in the figure agrees well with the data.  The
selection efficiency is $\epsilon\approx90$\% and the purity of the
samples is about $p\approx80$\%~\cite{l3290}.  The remaining fraction
of background from four-fermion events in the OPAL
analysis~\cite{OPALPR404} is 2.1\% at 161~GeV and 6.2\% at 207~GeV.  

The experimental corrections are typically \order{10}\% except at the 
edges of phase space.  They reach values of 50\% only for the LEP~2
samples because of the reduced selection efficiency of multi-jet final
states due to the additional cuts to suppress four-fermion events.

Some 3-jet observables have had QCD predictions in next-to-leading
order (NLO) combined with next-to-leading-log-approximation (NLLA)
predictions (\thr, \mh, \bt, \bw, \cp, \ytwothreed).  Most power
correction calculations are only available for these observables or a
subset thereof.  The final LEP results
are~\cite{l3290,aleph265,OPALPR404,delphi327} while many older
measurements are summarised in~\cite{colrun}.  Moments up to the
order~5 of these and several other observables are also
available~\cite{OPALPR404}.  The 4-jet observables \dpar, \bn, \ml,
\tmin\ and \ythreefourd\ were measured by OPAL and partially by 
L3~\cite{OPALPR404,l3290}.

\section{TRADITIONAL DMW STUDIES}

The power correction model of Dokshitzer, Marchesini and Webber (DMW)
extracts the structure of power correction terms from analysis of
infrared renormalon contributions~\cite{dokshitzer95a}.  Earlier
versions of the model are~\cite{webber94,dokshitzer95}.  The model
assumes that a non-perturbative strong coupling \asnp\ exists around
and below the Landau pole and that the quantity
$\anull(\mui)=1/\mui\int_0^{\mui}\asnp(k)\mathrm{d}k$ can be defined.
The value of \mui\ is chosen to be safely within the perturbative
region, usually $\mui=2$~GeV.  A study of the branching ratio of
hadronic to leptonic $\tau$ lepton decays as a function of the
invariant mass of the hadronic final state supports the assumption
that the physical strong coupling is finite and thus integrable at
low energy scales~\cite{brodsky03}, see below.

The main result for the effects of power corrections on distributions
$F(y)$ of the event shape observables \thr, \mh\ and \cp\ is that
the perturbative prediction $\fpt(y)$ is
shifted~\cite{dokshitzer97a,dokshitzer98b,dokshitzer99a}:
\begin{equation}
  F(y)= \fpt(y-\cy P)
\label{equ_npshift}
\end{equation}
where \cy\ is an observable dependent constant and $P\sim
M\mui/Q(\anull(\mui)-\as)$ is universal, i.e.\ independent of the
observable~\cite{dokshitzer98b}.  The factor $P$ contains the
$1/Q=1/\roots$ scaling and the so-called Milan-factor $M=1.49$ for
$\nf=3$ which takes two-loop effects into account.  The
non-perturbative parameter \anull\ is explicitly matched with the
perturbative strong coupling \as.  For the event shape observables
\bt\ and \bw\ the predictions are more involved and the shape of the
pQCD prediction is modified in addition to the
shift~\cite{dokshitzer99a}.  For mean values of \thr, \mh\ and \cp\
the prediction is:
\begin{equation}
  \momone{y}= \momone{y}_{\mathrm{PT}}+\cy P
\label{equ_pcmean}
\end{equation}
For \momone{\bt} and \momone{\bw} the predictions are also more
involved due to the modification of the shape of the distributions.

\begin{figure}[htb!]
\begin{tabular}{cc}
\includegraphics[width=0.49\textwidth,clip]{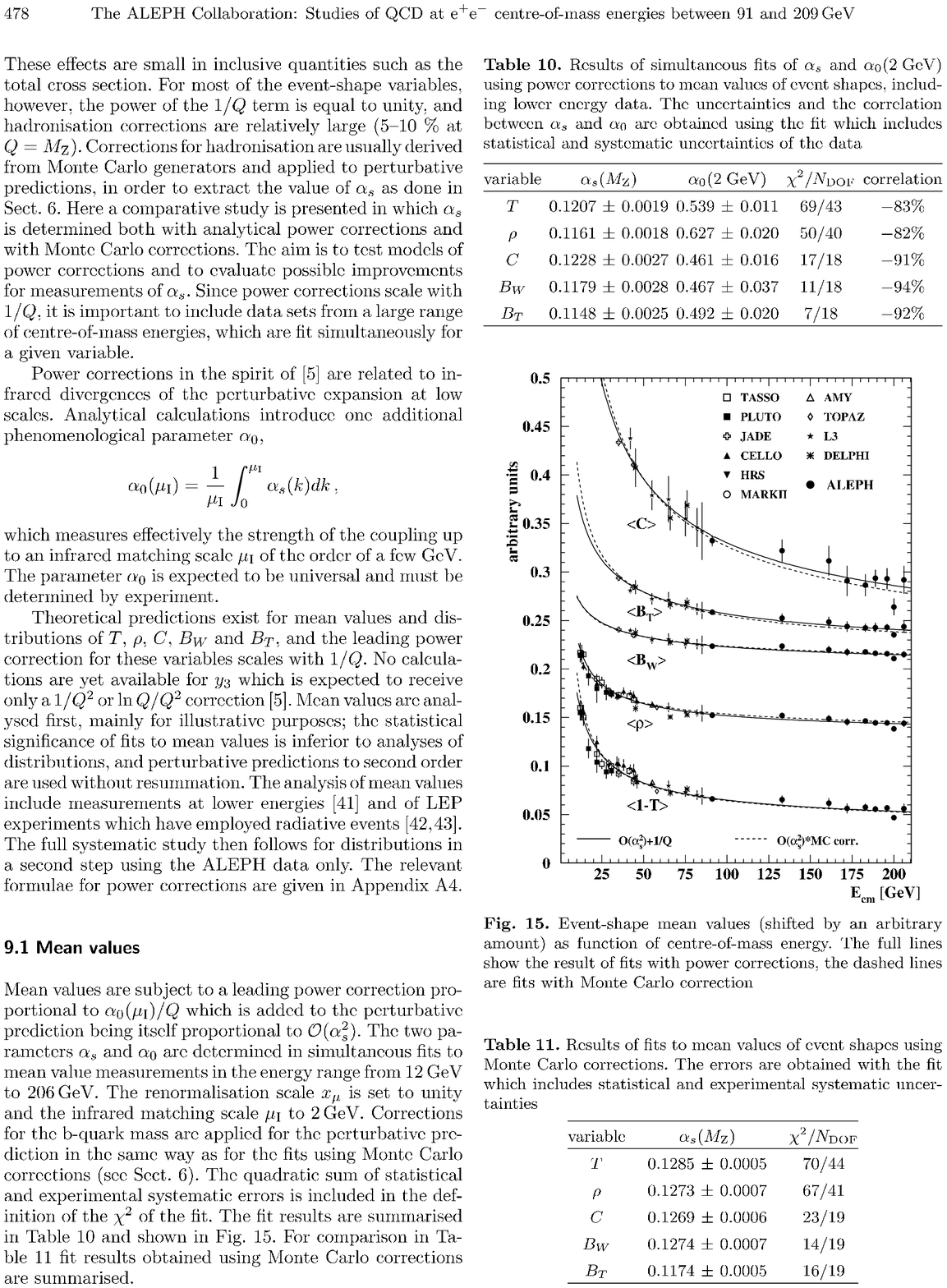} &
\includegraphics[width=0.49\textwidth,clip]{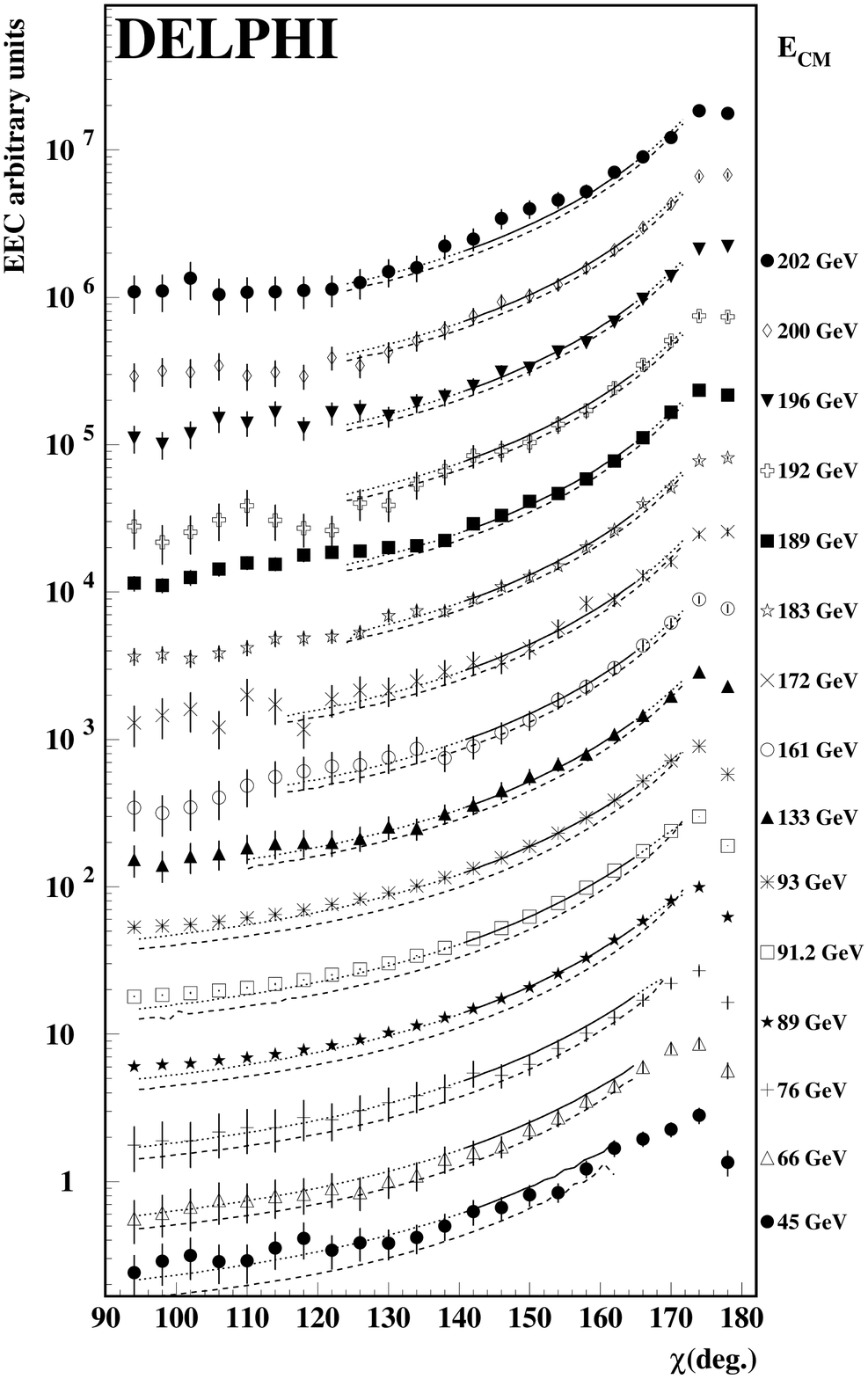} \\
\end{tabular}
\caption[ bla ]{ (left) Mean values of event shape observables as
indicated~\cite{aleph265}.  The solid lines are fits of NLO QCD
predictions combined with power corrections, the dashed lines present
fits of the same NLO QCD predictions with Monte Carlo based
hadronisation corrections.  (right) Measurements of EEC at various
\roots\ points as indicated~\cite{delphi296}.  The solid and dotted lines
show the fitted QCD prediction including power corrections using only
data points within the solid lines for the fit.  The dashed
lines present the same fit after subtraction of the power correction.}
\label{fig_momseec}
\end{figure}

Figure~\ref{fig_momseec} (left) shows the results of fits to data for
mean values of event shape observables measured over a large range of
cms energies~\cite{aleph265}.  The perturbative QCD prediction in NLO
$\momone{y}_{\mathrm{PT}}$ is combined with the power correction
calculations as explained above.  The combined prediction describes
the available data well within the uncertainties even at low cms
energies where non-perturbative effects are large.  The numerical results
are shown in table~\ref{tab_means}.  The dashed lines
show as a comparison the results of fits of the same NLO QCD
predictions with Monte Carlo based hadronisation corrections.  The
data are also well described but the difference in values of \asmz\
for the same observables is about 6\%~\cite{aleph265}. 

Figure~\ref{fig_momseec} (right) shows data of the event shape
observable EEC measured by DELPHI~\cite{delphi296} over a large range
of cms energies.  Superimposed are fits of the perturbative QCD
prediction in NLO+NLLA combined with power correction
calculations~\cite{dokshitzer99c}.  The distribution of EEC is
sensitive to a higher order non-perturbative parameter referred to as
\aone.  A fit with the three parameters \asmz, \anulltwo\ and
\aonetwo\ results in $\asmz=0.117\pm0.002$, $\anulltwo=0.48\pm0.05$
and $\aonetwo=0.00\pm0.04$.  The results for \asmz\ and \anulltwo\ are
consistent but the result for \aonetwo\ is not in agreement with an
expectation $\aonetwo=0.45$ from~\cite{dokshitzer99c}.

\begin{table}[!htb]
\begin{center}
\begin{tabular}{lrcccc}
\hline
\hline
 & Exp.        & ALEPH~\cite{aleph265} & DELPHI~\cite{delphi296} 
 & L3~\cite{l3290} & MPI~\cite{powcor} \\
\multicolumn{2}{r}{\roots\ range [GeV]} & 12(35)--206 & 12--202 & 41--206 
 & 12(35)--189 \\
\hline
 & \asmz         & $0.1207\pm0.0019$ & $0.1241\pm0.0034$ & $0.1164\pm0.0060$ 
 & $0.1217\pm0.0060$ \\
 1-T & \anulltwo & $0.539\pm0.011$   & $0.491\pm0.018$   & $0.518\pm0.059$ 
 & $0.528\pm0.064$ \\
 & \chisqd       & 69/43             & 27/41             & 18/14           
 & 50/41 \\
\hline
& \asmz          & $0.1161\pm0.0018$ & $0.1197\pm0.0122$ & $0.1051\pm0.0051$ 
 & $0.1165\pm0.0043$ \\
\mh, $\rho$ & \anulltwo & $0.627\pm0.020$   & $0.339\pm0.263$   
 & $0.421\pm0.037$ & $0.663\pm0.095$ \\
& \chisqd        & 50/40             & 10/15             & 13/14           
 & 24/35 \\
\hline
& \asmz          & $0.1148\pm0.0025$ & $0.1174\pm0.0029$ & $0.1163\pm0.0042$
 & $0.1205\pm0.0049$ \\
 \bt & \anulltwo & $0.492\pm0.020$   & $0.463\pm0.033$   & $0.449\pm0.054$   
 & $0.445\pm0.054$ \\
& \chisqd        & 7/18              & 9/23              & 9/14             
 & 24/28 \\
\hline
& \asmz         & $0.1179\pm0.0028$ & $0.1167\pm0.0019$ & $0.1169\pm0.0042$ 
 & $0.1178\pm0.0025$ \\
\bw & \anulltwo & $0.467\pm0.037$   & $0.438\pm0.049$   & $0.342\pm0.079$   
 & $0.425\pm0.097$ \\
& \chisqd       & 11/18             & 10/23             & 14/14             
 & 10/29 \\
\hline
& \asmz          & $0.1228\pm0.0027$ & $0.1222\pm0.0036$ & $0.1164\pm0.0047$ 
 & $0.1218\pm0.0059$ \\
 \cp & \anulltwo & $0.461\pm0.016$   & $0.444\pm0.022$   & $0.457\pm0.040$  
 & $0.461\pm0.048 $ \\
& \chisqd        & 17/18             & 12/23             & 12/14            
 & 18/26 \\
\hline
& \asmz          &  &    & $0.1046\pm0.0124$ &     \\
 \dpar & \anulltwo & $--$ & $--$ & $0.682\pm0.096$   & $--$ \\
& \chisqd        &  &    & 24/14             &     \\
\hline
\hline
\end{tabular}
\caption{Results for \asmz\ and \anulltwo\ from the analysis of mean
values in the DMW power correction model.  The errors consist of
statistical and experimental uncertainties only for ALEPH and DELPHI
and include also theoretical uncertainties for L3 and MPI.}
\label{tab_means}
\end{center}
\end{table}

\begin{figure}[htb!]
\includegraphics[width=0.75\textwidth]{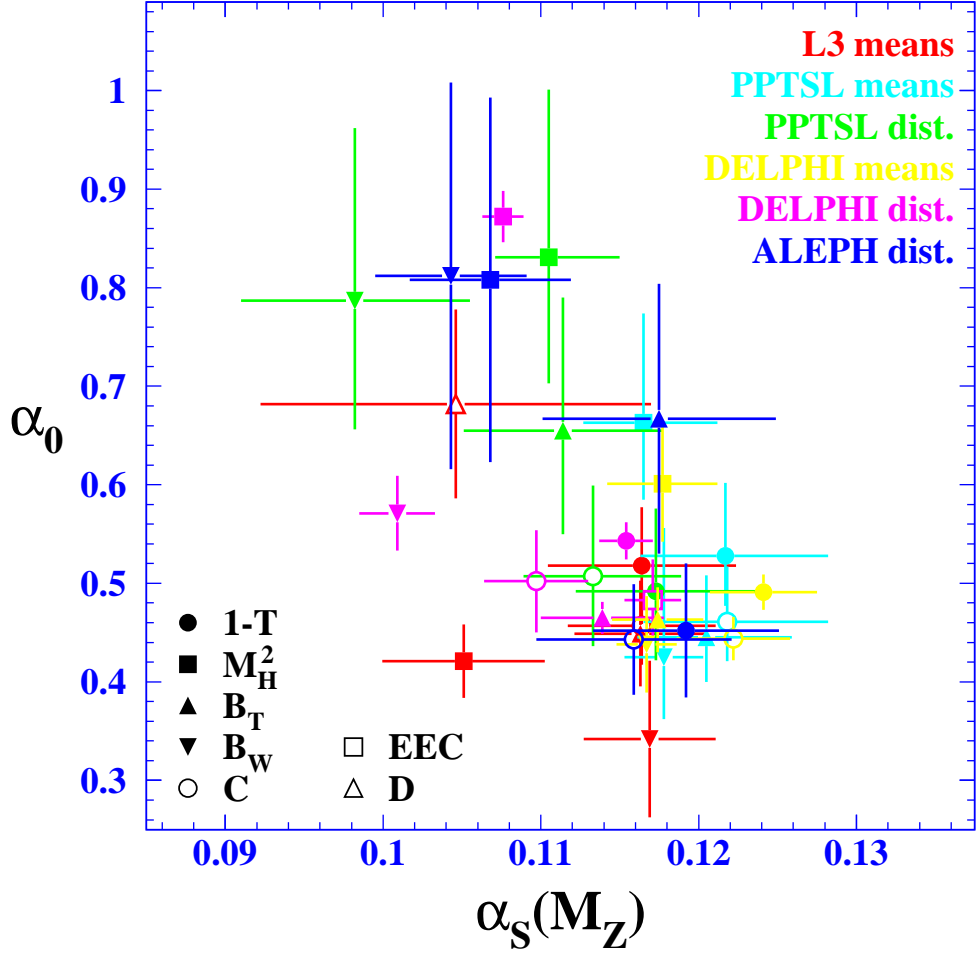}
\caption[ bla ]{ Summary of results from power correction 
studies~\cite{kluth06}.  Data are 
from~\cite{powcor,l3290,delphi296,aleph265}. }
\label{fig_allasa0}
\end{figure}

Tables~\ref{tab_means} and~\ref{tab_dists} summarise all recently
published results from studies of DMW power correction using mean
values and distributions of event shape observables.  The values for
\asmz\ and \anulltwo\ are generally consistent with each other except
for the following observations:
\begin{itemize}

\item The values for \anulltwo\ from \mh\ or $\rho$ are comparatively large.
  This is partially due to the effects of neglected hadron
  masses~\cite{salam01a}.  The results from DELPHI take hadron masses
  into account according to~\cite{salam01a} and their values for
  \anulltwo\ are lower.

\item The results for \anulltwo\ from distributions of \bt\ and \bw\ 
  from DELPHI appear inconsistent with the other results as noted
  already in~\cite{aleph265}.

\item The results for \anulltwo\ are consistent at the level of
  20 to 30\% but the errors including theoretical effects are
  partially smaller and are about 10\% in the case of the MPI analysis
  of mean values.  The Milan factor $M$ has a theoretical uncertainty
  of about 20\%~\cite{dokshitzer98b}.  However, it is not clear if
  this uncertainty should be reflected in differing values of
  \anulltwo\ from different observables.

\end{itemize}
The correlations between the two parameters \asmz\ and \anulltwo\ are
about $-90$\% from the fits and between about $-40\%$ and 0\% when
systematic effects are considered.  The results are also shown in
figure~\ref{fig_allasa0}.  The negative correlation between \asmz\ and
\anulltwo\ is visible in the figure.  

\begin{table}[!htb]
\begin{center}
\begin{tabular}{lrccc}
\hline
\hline
 & Exp.        & ALEPH~\cite{aleph265} & DELPHI~\cite{delphi296} 
 & MPI~\cite{powcor} \\
\multicolumn{2}{r}{\roots\ range [GeV]} & 91--206 & 45--202 & 14(35)--189 \\
\hline
 & \asmz &       $0.1192\pm0.0059$ & $0.1154\pm0.0017$ & $0.1173\pm0.0057$ \\
 1-T & \anulltwo & $0.452\pm0.068$   & $0.543\pm0.014$   & $0.492\pm0.077$   \\
 & \chisqd       & 73/47             & 291/180           & 172/263           \\
\hline
& \asmz          & $0.1068\pm0.0051$ & $0.1056\pm0.0007$ & $0.1105\pm0.0040$ \\
\mh, $\rho$ & \anulltwo & $0.808\pm0.185$   & $0.692\pm0.012$   & $0.831\pm0.149$   \\
& \chisqd        & 124/42            & 120/90            & 137/161           \\
\hline
& \asmz         & $0.1175\pm0.0074$ & $0.1139\pm0.0016$ & $0.1114\pm0.0063$ \\
\bt & \anulltwo & $0.667\pm0.137$   & $0.465\pm0.014$   & $0.655\pm0.120$   \\
& \chisqd       & 181/59            & 88/75             & 92/159            \\
\hline
& \asmz          & $0.1043\pm0.0048$ & $0.1009\pm0.0016$ & $0.0982\pm0.0073$ \\
 \bw & \anulltwo & $0.812\pm0.196$   & $0.571\pm0.031$   & $0.787\pm0.153$   \\
& \chisqd        & 76/47             & 106/90            & 96/132            \\
\hline
& \asmz          & $0.1159\pm0.0062$ & $0.1097\pm0.0032$ & $0.1133\pm0.0050$ \\
 \cp & \anulltwo & $0.443\pm0.056$   & $0.502\pm0.047$   & $0.507\pm0.082$   \\
& \chisqd        & 83/54             & 191/180           & 150/208           \\
\hline
& \asmz          &      & $0.1171\pm0.0018$ &      \\
 EEC & \anulltwo & $--$ & $0.483\pm0.041$   & $--$ \\
& \chisqd        &      & 53/90             &      \\
\hline
\hline
\end{tabular}
\caption{Results for \asmz\ and \anulltwo\ from the analysis of distributions
in the DMW power correction model. The errors consist of statistical
and experimental uncertainties only for ALEPH and DELPHI and include also
theoretical uncertainties for MPI. }
\label{tab_dists}
\end{center}
\end{table}

The values of \asmz\ are consistent with a recent determination of
\asmz\ from \epem\ annihilation data for event shape and jet
observables~\cite{kluth06}:
$\asmz=0.1202\pm0.0010\expt\pm0.0055\theo$.  The comparatively small
statistical and hadronisation uncertainties have been included in the
experimental and theoretical errors, respectively.  This result is
based on distributions corrected for hadronisation with Monte Carlo
models and can most directly be compared with the results from
distributions shown in table~\ref{tab_dists} where all values of
\asmz\ are lower.  The ALEPH analysis~\cite{aleph265} makes a direct
comparison of results for \asmz\ using DMW power corrections and Monte
Carlo hadronisation corrections and finds a difference of about 9\%.
This difference is larger than the commonly quoted total error of
about 5~to 6\% on determinations of \asmz\ from event shape or jet
observables.

\section{EXTENDED DMW FITS}

The MPI analysis of distributions of \bw~\cite{powcor} finds large
values for \chisqd\ when $\roots<\mz$ indicating a beginning failure
of the available calculations to describe the low energy data.  For
distributions of \ytwothreed\ a similar analysis~\cite{pedrophd}
finds large values of \chisqd\ when data at low \roots\ are included.

\begin{figure}[htb!]
\begin{tabular}{cc}
\includegraphics[width=0.49\textwidth]{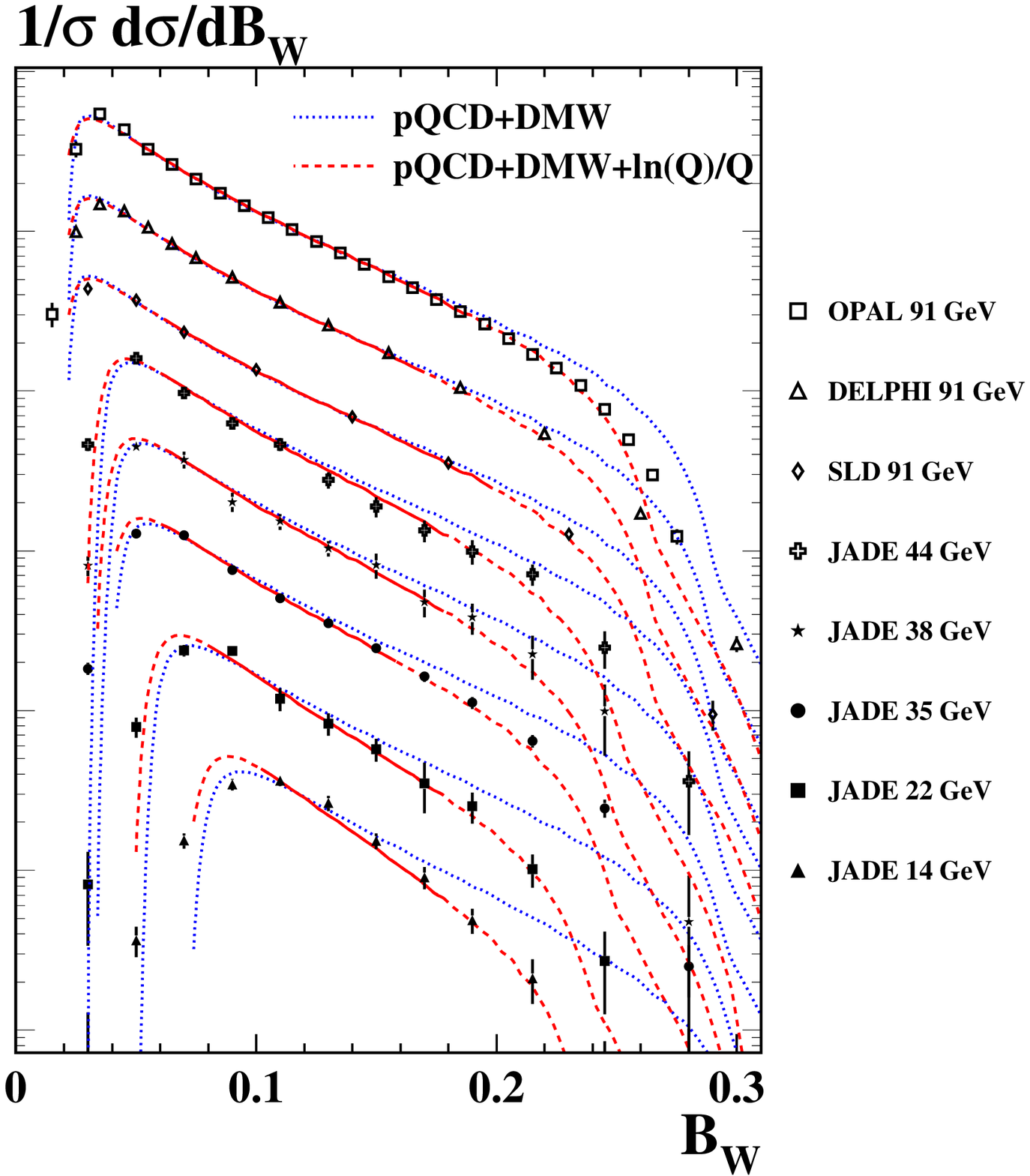} &
\includegraphics[width=0.49\textwidth]{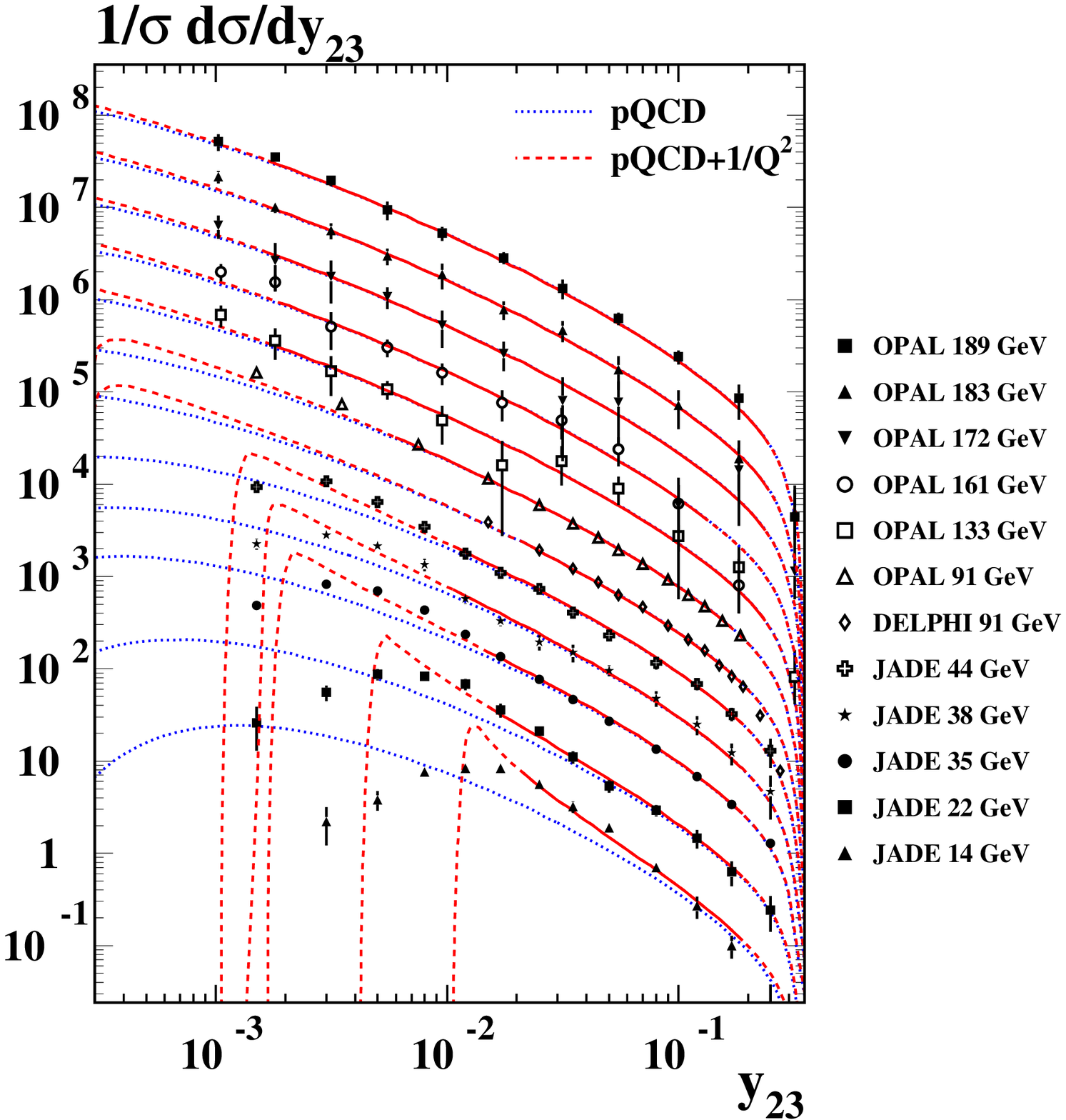} \\
\end{tabular}
\caption[ bla ]{ (left) Distributions of \bw\ measured at
$14\le\roots\le91$~GeV as shown on the figure~\cite{pedrophd}.
Superimposed are fits perturbative QCD combined with power corrections
with (solid and dashed lines) and without (dotted lines) additional
terms as explained in the text. (right) Distributions of \ytwothreed\
measured at $14\le\roots\le189$~GeV as shown on the
figure~\cite{pedrophd}.  Superimposed are fits of pure perturbative
QCD (solid and dashed lines) or perturbative QCD combined with a power
correction term as explained in the text.}
\label{fig_bwy23ext}
\end{figure}

Figure~\ref{fig_bwy23ext} (left) shows results of simultaneous fits of
\asmz\ and \anulltwo\ to data for distributions of \bw\ at
$14\le\roots\le91$~GeV~\cite{pedrophd}.  The solid lines indicate
results of extended fits where an additional term $A_{11}\ln(Q)/Q$,
$Q=\roots$, with a new free parameter $A_{11}$ has been added to the
DMW prediction.  The fit describes the data well and yields
$\chisqd=22/23$.  The same fit without the additional term is shown by
the dotted lines; the fit does not describe the data at low \roots\
well and $\chisqd=66/24$ is found.  One thus has evidence for
additional contributions to this observables at low \roots\ probably
behaving like $\ln(Q)/Q$.

Figure~\ref{fig_bwy23ext} (right) shows results of simultaneous fits of
\asmz\ and \anulltwo\ to data for distributions of \ytwothreed\ at
$14\le\roots\le189$~GeV.  The solid lines show fits of the pure
perturbative QCD prediction with an additional term of the form
$A_{20}/Q^2$ inspired by~\cite{dokshitzer95a}.  The fit describes the
data at all cms energies and $\chisqd=71/106$ is found.  The dotted
lines present the results of the same fits without the additional
power correction term.  The fit does not describe well the data at low
cms energies, in particular at $\roots=14$ and 22~GeV.  One thus
has evidence for additional terms probably behaving like $1/Q^2$
at low \roots.  At large cms energies the two types of fit are 
very similar.  

\section{EVIDENCE FOR SOFT FREEZING}

One of the main assumptions of the DMW model of power corrections is,
simply speaking, that the physical running strong coupling \asnp\
remains finite around and below the Landau pole, where the
perturbative evolution of the strong coupling according to the
renormalisation group brakes down.  With this assumption quantities
like $\anull(\mui)=1/\mui\int_0^{\mui}\asnp(k)\mathrm{d}k$ can be
defined.

Using hadronic decays of $\tau$ leptons this question can be studied
experimentally.  The $\tau$ lepton decays weakly via an intermediate
(virtual) \wplus\ or \wminus\ boson.  When the W boson decays to
quarks QCD effects have to be taken into account.  The mass of
the virtual W bosons $\mw\le m_{\tau}=1.777$~GeV sets the scale for these
effects which are rather large.  The spectrum of invariant masses of
hadronic final states from $\tau$ lepton decays provides a direct
experimental measurement of the scale given by the virtual \mw.

\begin{figure}[htb!]
\begin{tabular}{cc}
\includegraphics[width=0.49\textwidth]{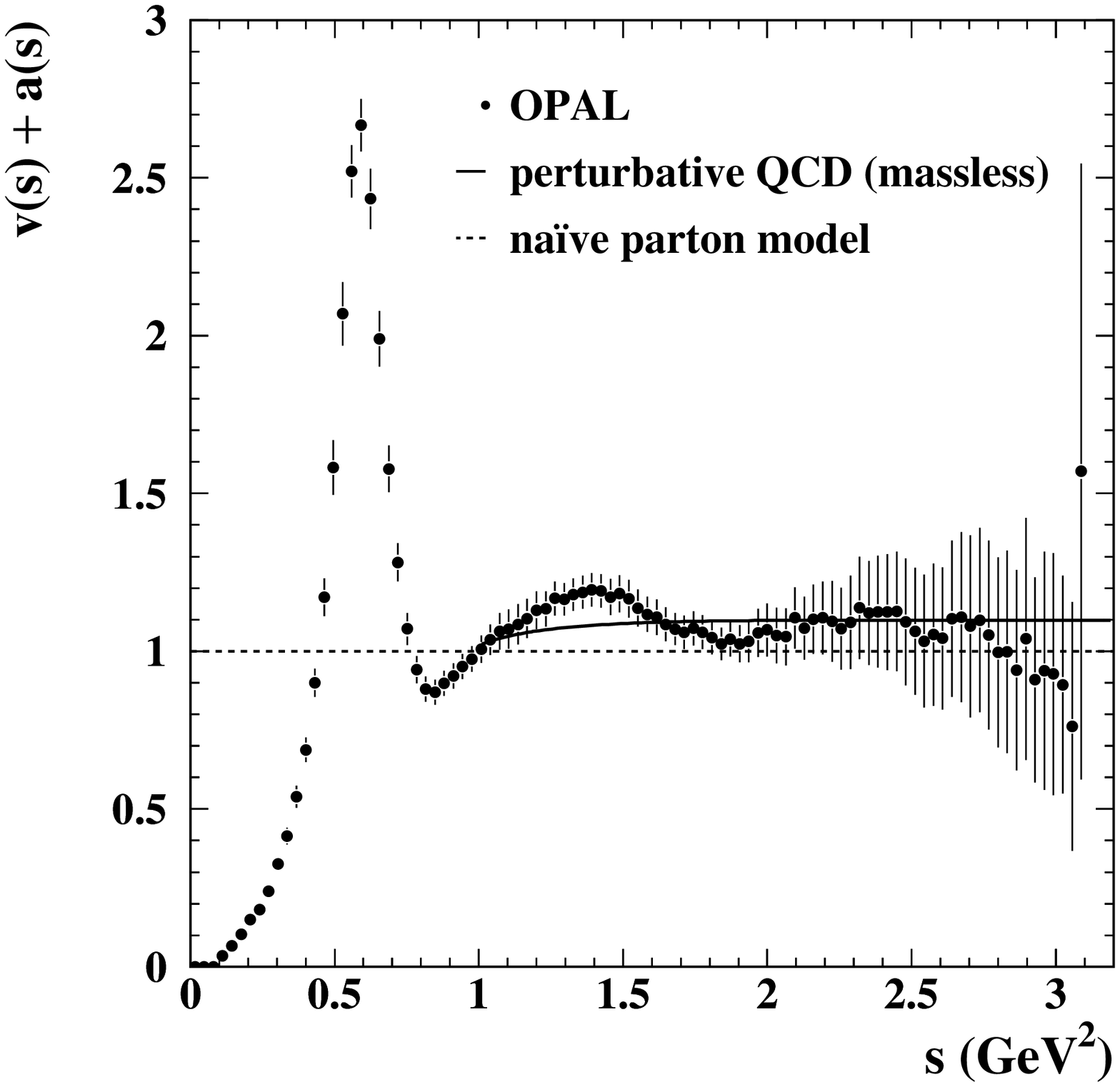} &
\includegraphics[width=0.49\textwidth]{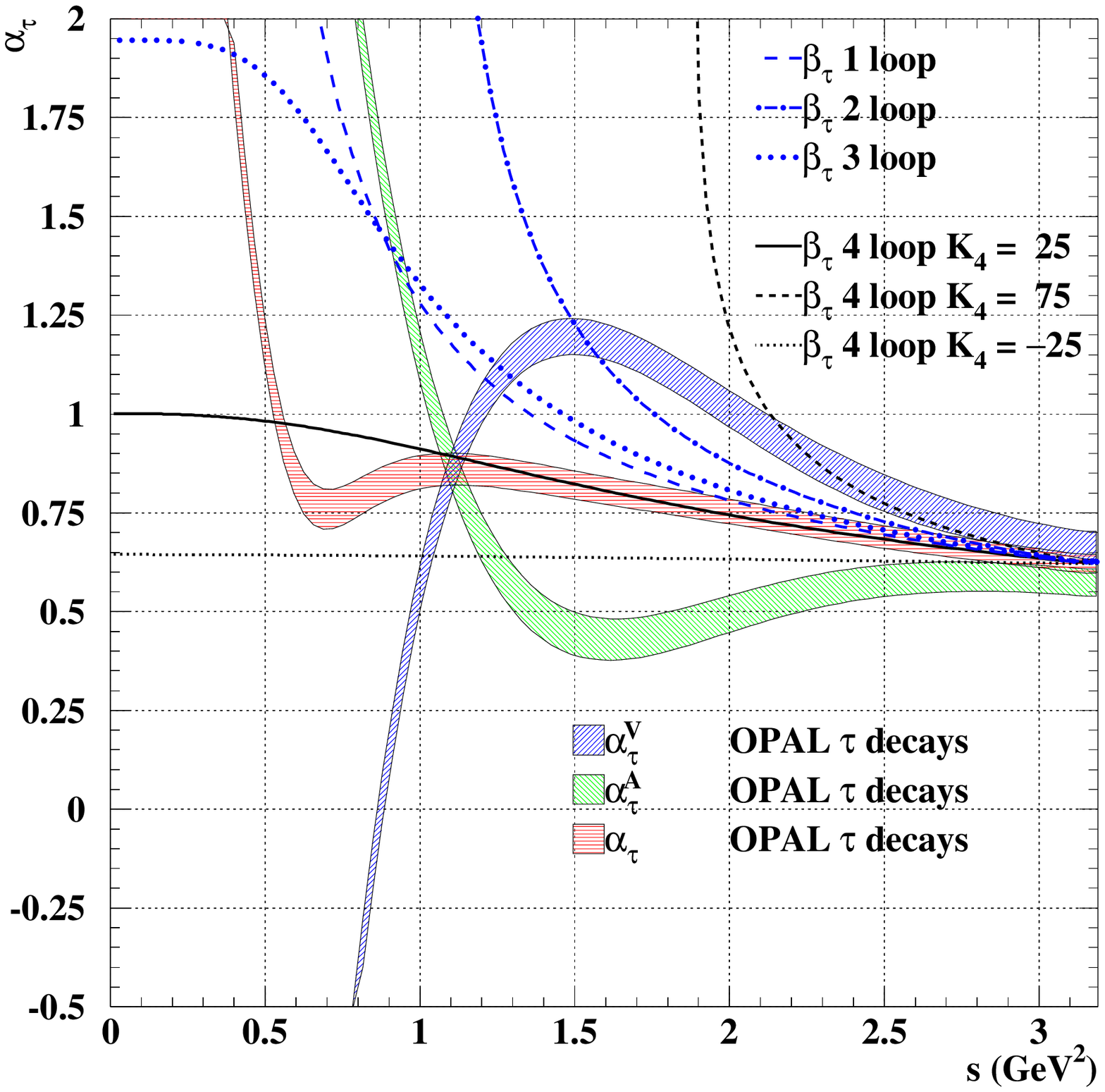} \\
\end{tabular}
\caption[ bla ]{ (left) Spectrum of invariant masses of $\tau$
lepton decays into non-strange hadronic final states~\cite{OPALPR246}.
The solid and the dashed line show predictions as indicated on the
figure.  (right) The effective charge \atau\ determined from hadronic
$\tau$ decays as explained in the text as a function of
$s$~\cite{brodsky03}.  The lines show varios predictions
for $\atau(s)$ as explained in the text. }
\label{fig_softfreeze}
\end{figure}

Figure~\ref{fig_softfreeze} {left} shows a measurement of the
inclusive spectrum of invariant masses of non-strange hadronic final
states $s=m_h^2$ by OPAL~\cite{OPALPR246}.  The quantity
$v(s)+a(s)=\ddel R_{\tau}/\ddel s$ with
$R_{\tau}(s)=\Gamma(\tau\rightarrow\mathrm{h}\nu_{\tau})/
\Gamma(\tau\rightarrow\ell\nu_{\ell}\nu_{\tau})$.  The dashed line
shows the ``parton model prediction'' for three colours.  The solid
line shows the perturbative QCD prediction for massless quarks.  The
line describes the correlated data points well; this implies that
non-perturbative corrections are small for the inclusive spectrum.

The analysis~\cite{brodsky03} defines an effective charge $\atau(s)$
via $R_{\tau}(s)=R_{\tau}^0(1+\atau(s)/\pi)$.  The relation between
$\atau(s)$ and $\as(s)$ is known perturbatively.
Figure~\ref{fig_softfreeze} (right) shows the variable $\atau(s)$ as a
function of $s$ extracted from the data (shaded bands) and using
different predictions (lines).  The band from analysis of the
inclusive spectrum labelled ``\atau'' stays fairly flat down to
$s\simeq 1$~GeV$^2$ and then rises steeply due to the $\rho$
resonance.  The other bands are from analysis of spectra for even or
odd number of pions in the final state and have strong
non-perturbative effects.  The solid line labelled ``$\beta_{\tau}$ 4
loop $K_4=25$'' corresponds to the most complete QCD prediction for
the running of the effective charge including an estimate of 4th order
term.  This prediction describes the data well down to $s\simeq
1$~GeV$^2$ and stays finite at $s=0$.  One also observes that the NNLO
prediction labelled ``$\beta_{\tau}$ 3 loop'' also describes the data
at medium values of $s$ and is finite at $s=0$.  It is thus possible
to consistently describe data for hadronic $\tau$ decays down to
$s\simeq1$~GeV$^2$ with an effective coupling which stays finite at
$s=0$.  These observations are consistent with earlier
work~\cite{howe02,howe03}.

\section{SUMMARY AND OUTLOOK}

In this review several topics were not covered due to a lack of
results.  One issue is the study of 4-jet event shape observables in
the planar limit.  The theoretical aspects and preliminary
experimental results are discussed in contributions B001 and B002 to
this conference.  However, there are no published experimental studies
which could be subject of this review, but hopefully there will
be results on this important subject in the near future.  Other
models for the description of soft corrections to hard QCD processes
such as shape functions, the single-dressed-gluon approximation (SDG)
are subject of contributions to this conference but there are no
detailed experimental studies.  From an experimental point of
view the DMW model is attractive, because it makes strong predictions
about relations between observables and has only universal free
parameters.  Also, predictions actually exist for several observables
such that direct comparisons are possible.

The power corrections in the DMW model have been studied by several
experimental groups as presented in this review.  The picture is generally 
consistent with only a few notes:
\begin{itemize}

\item The values for \anulltwo\ from different observables are consistent
  within 20 to 30\%.  However, it has to be understood if this is a
  variation one expects from the theoretical uncertainties in the
  calculations.  It is often claimed that the theoretical uncertainty
  on the Milan factor $M$ of about 20\% due to missing higher order
  terms could explain the scatter of \anulltwo\ results.  However,
  since $M$ is a universal factor it is not clear if it can explain
  differences between different observables.

\item The values for \asmz\ from different observables are consistent
  with each other within the uncertainties of the perturbative QCD
  calculations as expected.  However, direct comparison of the results
  from power correction analyses and analyses with hadronisation
  corrections derived from Monte Carlo simulation show a systematic
  bias for \asmz\ of about 6\% (mean values) or 9\% (distributions).
  This bias is comparable to or larger than the total uncertainties
  quoted for determinations of \asmz\ from event shape or jet
  observables in \epem\ annihilation.  The bias is significantly
  larger than the hadronisation model uncertainties of less than 2\%
  quoted for \asmz\ determinations using LEP data~\cite{kluth06}.

\end{itemize}

In the future the studies of power corrections with \epem\ annihilation
data would profit from further analyses using the JADE data
in the intermediate energy region $14\le\roots\le44$~GeV.  In
particular the EEC and 4-jet observables with or without requiring
the planar limit would be attractive subjects.

The impact of NNLO perturbative QCD calculations for 3-jet observables
could also be significant.  Firstly, the power correction calculations
will have to adapted to the improved predictions. Secondly, with there
is potential for high precision measurements of \asmz\ with reduced
theoretical uncertainties.  With selected observables with suppressed
power corrections such as \ytwothreed\ one could obtain the matching
reduced hadronisation uncertainties.

Other approaches to understand soft corrections in terms of QCD should
be studied in more detail.  The connection with RS optimisation
discussed in contributions T001 and T002 is also an interesting new
area of research.


\end{document}